\documentclass[11pt]{article}

\usepackage{algorithm}
\usepackage{algorithmic}
\usepackage{amsmath,amssymb}
\usepackage[colorlinks]{hyperref}
\usepackage[margin=1.0in]{geometry}
\usepackage[numbers]{natbib}
\newtheorem{theorem}{Theorem}[section]
\newtheorem{lemma}[theorem]{Lemma}

\newtheorem{counter-example}[theorem]{Counter example}

\newtheorem{open question}[theorem]{Open question}

\newcommand{\ignore}[1]{}

\newcommand{\cd}{{\cal D}}

\newcommand{\cf}{{\cal F}}

\newcommand{\cy}{{\cal Y}}

\DeclareMathOperator*{\erf}{erf}

\DeclareMathOperator*{\sign}{sign}

\newcommand{\reals}{{\mathbb R}}

\newcommand{\POL}{\mathrm{POL}}

\newcommand{\proof}{{\par\noindent {\bf Proof}\space\space}}
\newcommand{\proofbox}{\begin{flushright}$\Box$\end{flushright}}

\DeclareMathOperator{\Err}{Err}

\DeclareMathOperator{\poly}{poly}

\newcommand{\opt}{{\mathrm{opt}}}

\newcommand{\inner}[1]{\langle #1 \rangle}

\title{A PTAS for Agnostically Learning Halfspaces}
\author{Amit Daniely \thanks{Department of Mathematics, Hebrew
                 University, Jerusalem 91904, Israel.  amit.daniely@mail.huji.ac.il}}

\begin{document}
\maketitle
\setcounter{page}{0}

\thispagestyle{empty}
\maketitle

\begin{abstract} 
We present a PTAS for agnostically learning halfspaces w.r.t. the uniform distribution on the $d$ dimensional sphere. Namely, we show that for every $\mu>0$ there is an algorithm that runs in time $\poly\left(d,\frac{1}{\epsilon}\right)$, and is guaranteed to return a classifier with error at most $(1+\mu)\opt+\epsilon$, where $\opt$ is the error of the best halfspace classifier. This improves on Awasthi, Balcan and Long \cite{AwasthiBalcanLong14} who showed an algorithm with an (unspecified) constant approximation ratio.
Our algorithm combines the classical technique of polynomial regression (e.g. \cite{LinialMaNi89, KalaiKlMaSe05}), together with the new localization technique of \cite{AwasthiBalcanLong14}.
\end{abstract}

\newpage

\section{Introduction}

In the problem of agnostically learning halfspaces, the learner is given an access to examples drawn from a distribution $\cd$ on $\mathbb{R}^d\times \{\pm 1\}$ and an accuracy parameter $\epsilon>0$. It is required to output\footnote{Throughout, we require our algorithms to succeed with a constant probability (that can be standardly amplified by repetition).} a classifier $h:\mathbb{R}^d\to \{\pm 1\}$ whose error, $\Err_{\cd}(h):=\Pr_{(x,y)\sim\cd}\left(h(x)\ne y\right)$, is at most\footnote{Note that $\opt$ might be $>0$, namely, we consider the ``agnostic PAC learning" model \cite{KearnsScSe94}.} $\opt+\epsilon$. Here, $\opt$ is the error of the best classifier of the from $h_w(x)=\sign(\inner{w,x})$. The learner is {\em efficient} if it runs in time $\poly\left(d,\frac{1}{\epsilon}\right)$. We note that we consider the general, {\em improper}, setting where the learner have the freedom to return a hypothesis that is not a halfspace classifier.

Halfspaces are extremely popular in practical applications, and have been extensively studied in Machine Learning, Statistics and Theoretical Computer Science (see section \ref{sec:related_work}).
Unfortunately, from a worst case perspective, the problem seems very hard: Best known efficient algorithms have a terrible approximation ratio of $\tilde{\Omega}(d)$.
In the case of {\em proper learning}, where the output hypothesis must be a halfspace, agnostic learning is known to be $\mathcal{NP}$-hard. Even learning with a constant {\em approximation ratio}, where the returned classifier should have error $\le \alpha\cdot\opt+\epsilon$, is $\mathcal{NP}$-hard. In fact, even approximation ratio of $2^{\log^{0.99}(d)}$ is $\mathcal{NP}$-hard.
In the general (improper) case, agnostic learning of halfspaces, and even agnostic learning with an approximation ratio of $2^{\log^{0.99}(d)}$, have been showed hard under various complexity assumptions (see section \ref{sec:related_work}).
In light of that, it is just natural to consider agnostic learning under various
restrictions on the distribution $\cd$. A very natural and widely studied such restriction \cite{KlivansLoSe09, klivans2002learning, AwasthiBalcanLong14, KalaiKlMaSe05} is that the marginal distribution, $\cd_{\mathbb R^d}$, is uniform on the sphere $S^{d-1}$.

Even under the uniform distribution, no efficient algorithms are known, and there is also an evidence that the problem is hard \cite{klivans2014embedding}. This lead researchers to consider {\em approximation algorithms}. The first approximation guarantee is due to \cite{KalaiKlMaSe05}, who showed an efficient regression based algorithm with approximation ratio of $\alpha=O\left(\sqrt{\log\left(\frac{1}{\opt}\right)}\right)$. In an exciting recent work, \cite{AwasthiBalcanLong14} introduced a new algorithmic technique, called {\em localization}, and showed an efficient algorithm with an unspecified constant approximation ratio. In this paper, we advance this line of work further, and show a Polynomial Time Approximation Scheme (PTAS). Namely, we show:
\begin{theorem}[main]\label{thm:main}
For every $\mu>0$, there is an efficient algorithm for agnostically learning halfspaces under the uniform distribution with an approximation ratio of $(1+\mu)$.
\end{theorem}
As noted above, \cite{klivans2014embedding} showed that under a certain complexity assumption (hardness of learning sparse parity), there are no exact efficient algorithms (i.e., with approximation ratio $\alpha=1$). In that case, our result is optimal.

{\bf Label Complexity:} Our algorithm naturally fits to the {\em active learning} (e.g. \cite{settles2010active}) setting.  Often, a label is much more expensive than an example (e.g., when applying learning methods in biology, it might be the case that we have to make an experiment in order to get a label). It is therefore useful that algorithms will make economical use of labels. Our algorithm naturally have such property, as its {\em label complexity} (i.e., the number of labels it needs to see) is poly-logarithmic in $\frac{1}{\opt}$ (see theorem \ref{thm:main_detailed} for a more detailed statement).

{\bf Interpolation between approximation and exact algorithms:} A more precise statement of our result is that there exists an algorithm with runtime $\poly\left(d^{\frac{\log^{3}\left(\frac{1}{\mu}\right)}{\mu^2}},\frac{1}{\epsilon}\right)$ that is guaranteed to return a classifier with error at most $(1+\mu)\opt+\epsilon$ for every $0<\mu,\epsilon\le 1$. Taking $\mu$ up to $\frac{\epsilon}{2}$ and replacing $\epsilon$ with $\frac{\epsilon}{2}$, the error bound is $\left(1+\frac{\epsilon}{2}\right)\opt+\frac{\epsilon}{2}\le\opt+\epsilon$. Hence, we get an exact algorithm. The running time is $\poly\left(d^{\frac{\log^{3}\left(\frac{1}{\epsilon}\right)}{\epsilon^2}}\right)$, which almost matches the current state of the art -- $\poly\left(d^{\frac{1}{\epsilon^2}}\right)$ \cite{KalaiKlMaSe05,diakonikolas2010bounded2}.

{\bf Open questions:} Obvious open questions are to extend our results to more distributions (uniform on $\{\pm 1\}^d$, permutation-invariant, product, log-concave, \ldots) and more problems (learning intersection of halfspaces, functions of halfspaces, \ldots). In addition, as opposed to previous approximation algorithms \cite{AwasthiBalcanLong14,KalaiKlMaSe05}, our algorithm  does not always return a halfspace classifier. A natural open question is therefore to find a {\em proper} PTAS.

\subsection{Algorithmic Components, The PTAS, and Proof Outline}
Our algorithm and its analysis  build on and combine various algorithmic and proof techniques that were previously used for learning halfspaces. This includes regression based algorithms (e.g. \cite{ShalevShSr11,KalaiKlMaSe05}), polynomial approximations of the sign function (e.g. \cite{ShalevShSr11,KalaiKlMaSe05, diakonikolas2010bounded,diakonikolas2010bounded2}) and localization techniques \cite{AwasthiBalcanLong14}. In this section we outline these techniques and the way we use them. Then, we present our PTAS, state its properties (theorem \ref{thm:main_detailed}), and describe the course of the proof. The full proof is in sections \ref{sec:proof} and \ref{sec:sign_appr}.

\subsubsection{Some preliminaries}

{\bf Noise tolerance} is a measure to evaluate the performance of learning algorithms, that is essentially equivalent to the approximation ratio. Yet, we find it slightly more convenient for the technical exposition.
We say that a learning algorithm tolerates noise rate of $0<f(\eta)<\eta$ (w.r.t. halfspaces) if, when running on input $0<\eta<1$, it guaranteed to return a hypothesis with error $\le \eta$, provided that $\opt\le f(\eta)$. We say that such an algorithm is {\em efficient} if it runs in time $\poly\left(d,\frac{1}{\eta}\right)$.
We note that given a learning algorithm that tolerates noise rate of $\frac{\eta}{\alpha}$, for some $\alpha>1$, it is not hard to construct an algorithm with approximation ratio of $\alpha$, and the running time grows only by a factor of $\poly\left(\frac{1}{\epsilon}\right)$: Indeed, in order to return a hypothesis with error $\le \alpha\cdot\opt+\epsilon$, we can run the algorithm with $\alpha\cdot\opt \le \eta\le \alpha\cdot\opt+\epsilon$. We can find such an $\eta$ by trying $\eta=k \epsilon$ for $k=1,2,\ldots,\left\lceil\frac{1}{\epsilon}\right\rceil$.
\\
{\bf Notation.} Let $\cd$ be a distribution on a space $X$. For $Y\subset X$ we denote by $\cd|_Y$ the restriction of $\cd$ to $\cy$. If $\cd$ is a distribution on $X\times \{\pm 1\}$ we denote by $\cd_X$ the marginal distribution on $X$. If $\cd$ is a distribution on $S^{d-1}$ (resp. $S^{d-1}\times \{\pm 1\}$) and $w\in S^{d-1}$, we define the {\em projection of $\cd$ on $w$} as follows: If $x\sim\cd$ (resp. $(x,y)\sim \cd$) then $\cd_w$ is the distribution (on $[-1,1]$) of the random variable $\inner{w,x}$.
For a distribution $\cd$ on a space $X$ and a function $f:S^{d-1}\to \mathbb{R}$, we denote $\|f\|_{p,\cd}=\left(\mathbb{E}_{x\sim\cd}|f(x)|^p\right)^{\frac{1}{p}}$. 
We will sometimes abuse notation and use $\|f\|_{p,\cd}$ instead of $\|f\|_{p,\cd_{S^{d-1}}}$ even when $\cd$ is a distribution on $S^{d-1}\times\{\pm 1\}$.
We denote by $\theta(w,w^*)=\cos^{-1}(\inner{w,w^*})$ the angle between two vectors $w,w^*\in S^{d-1}$. We will frequently use the fact that for uniform $x\in S^{d-1}$ we have $\Pr\left(h_{w^*}(x)\ne h_w(x)\right)=\frac{\theta(w,w^*)}{\pi}$. We denote by $\POL_{r,d}$ the space of $d$-variate polynomials of degree $\le r$. For $w\in S^{d-1}$ and $\gamma>0$ we let $T_{d,\gamma}(w):=\{u\in S^{d-1}: |\inner{w,u}|\le \gamma\}$.

\subsubsection{Polynomial $\ell_1$-regression for classification}
The output of a classification algorithm is a (description of a) hypothesis $h:S^{d-1}\to \{\pm 1\}$. Often, the returned hypothesis is of the form $h(x)=\sign(f(x))$, for some real valued function $f:S^{d-1}\to\mathbb R$. To conveniently dealing with such hypotheses, we introduce some terminology.
We denote the standard (zero-one) loss of $f$ by $\Err_{\cd}(f)=\Err_{\cd}\left(\sign(f)\right)$. We also consider the {\em $\ell_1$-loss}, $\Err_{\cd,1}(f)=\mathbb{E}_{(x,y)\sim\cd}|f(x)-y|$.
We note that for $f:S^{d-1}\to\mathbb{R}$, since $\frac{|\sign(z)-1|}{2}\le |z-1|$ for all $z$, we have
\begin{eqnarray*}
\Err_{\cd}(f)&=&\mathbb{E}_{(x,y)\sim\cd}\frac{|\sign(yf(x))-1|}{2}
\\
&\le & \mathbb{E}_{(x,y)\sim\cd}|yf(x)-1|
\\
&= & \mathbb{E}_{(x,y)\sim\cd}|f(x)-y| = \Err_{\cd,1}(f)
\end{eqnarray*}
Thus, by finding $f$ with small $\ell_1$-error we can find a good classifier. The motivation for moving from the 0-1 loss to the $\ell_1$ loss is the convexity of the $\ell_1$ loss, which enables the use of convex optimization. Concretely, for ``nice enough" convex set, $\cf$, of functions from $S^{d-1}$ to $\mathbb R$, it is possible to efficiently find (both in terms of number of examples and time) $f\in \cf$ with $\ell_1$ error almost as small as $\min_{f\in \cf}\Err_{\cd,1}(f)$.
Now, for a classifier $h:S^{d-1}\to\{\pm 1\}$ we have 
\begin{eqnarray}\label{eq:l_1_err_bounded_by_0_1}
\Err_{\cd}(f)\le \Err_{\cd,1}(f)&= & \mathbb{E}_{(x,y)\sim\cd}|f(x)-y|
\nonumber \\
&\le & \mathbb{E}_{(x,y)\sim\cd}|f(x)-h(x)|+\mathbb{E}_{(x,y)\sim\cd}|h(x)-y|
\\ \nonumber
&= & \|f-h\|_{1,\cd}+2\Err(h)
\end{eqnarray}
Thus, if we minimize the $\ell_1$-loss over a collection of functions that is large enough to contain a good $\ell_1$-approximation of the best halfspace classifier, we can find a function whose $\ell_1$-error, and therefore also the 0-1 error, is almost as good as the 0-1 error of the best halfspace classifier.
Methods that follow the above spirit have been extensively studied in computational learning theory. Concretely, \cite{KalaiKlMaSe05} suggested the following algorithm: First, find $P\in \POL_{r,d}$ that minimizes the empirical $\ell_1$-error on the given sample\footnote{I.e., if the sample is $(x_1,y_1),\ldots,(x_m,y_m)\in S^{d-1}\times\{\pm 1\}$, find $P\in \POL_{r,d}$ that minimizes $\frac{1}{m}\sum_{i=1}^m|P(x_i)-y_i|$.}. Then, find a classifier that makes the least number of errors on the given sample, among all classifiers of the form $x\mapsto\sign\left(P(x)-a\right)$ for $a\in \mathbb R$. We note that the second step is required in order to overcome the factor of $2$ in equation (\ref{eq:l_1_err_bounded_by_0_1}). They used that algorithm to show:
\begin{theorem}\cite{KalaiKlMaSe05}\label{thm:regression}
There is an algorithm with runtime $\poly\left(d^r,\frac{1}{\epsilon}\right)$ such that, for every distribution $\cd$ on $S^{d-1}\times\{\pm 1\}$ and every $h:S^{d-1}\to\{\pm 1\}$, it returns $P\in \mathbf{POL}_{r,d}$ with $\Err_{\cd}(P)\le \Err_{\cd}(h)+\min_{P'\in \mathrm{POL}_{r,d}}\|h-P'\|_{1,\cd}+\epsilon$.
\end{theorem}
\subsubsection{Learning halfspaces using sign approximations}\label{sec:sign_appr_intro}
To use theorem \ref{thm:regression} for learning halfspaces, we need to prove the existence of low degree polynomials $P$ such that $\|h-P\|_{1,\cd}$
is small, where $h$ is a halfspace classifier.
As explained below, this is naturally done by approximating the {\em sign function}, \mbox{$\sign(x)=\begin{cases}1 & x>0\\-1 & x \le 0\end{cases}$}, with respect to an appropriate proximity measure.

Suppose that $w^*\in S^{d-1}$ 
defines the optimal halfspace and let $\cd_{w^*}$ be the projection of $\cd$ 
on $w^*$. For a univariate polynomial $p\in \POL_{r,1}$, consider the $d$-variate polynomial $P\in\POL_{r,d}$ given by $P(x)=p(\inner{w^*,x})$. We have
\begin{eqnarray}\label{eq:appr_half_by_appr_sign}
\|P-h_{w^*}\|_{1,\cd} &=& \mathrm{E}_{x\sim\cd_{S^{d-1}}}[|p(\inner{w^*,x})-\sign(\inner{w^*,x})|]
\nonumber \\
&=& \mathrm{E}_{x\sim\cd_{w^*}}[|p(x)-\sign(x)|]
\\
&=& \|p-\sign\|_{1,\cd_{w^*}} \nonumber
\end{eqnarray}
Therefore, in order to find a good $\ell_1$ approximation for $h_{w^*}$ w.r.t. $\cd$, we can find a good $\ell_1$ approximation for $\sign$ w.r.t. $\cd_{w^{*}}$.

Approximating the sign function is a central component in many papers about halfspaces \cite{BirnbaumSh12, diakonikolas2010bounded, diakonikolas2010bounded, KalaiKlMaSe05, ShalevShSr11}. These papers needed to find approximation of the sign function w.r.t. relatively well studied proximity measures, such as the $\ell_\infty$ norm, or the $\ell_1$ and $\ell_2$ norms w.r.t. the Gaussian distribution. Therefore, some of these papers used basis expansion methods (Fourier, Hermite, Chebyshev, \ldots). In this paper we need to find $\ell_1$ approximation w.r.t. messier distributions. Therefore, we use a somewhat more flexible approach, similar to the one used in \cite{diakonikolas2010bounded}.
We rely on techniques from approximation theory \cite{davis1975interpolation}. In particular, our main tool for constructing polynomials will be the celebrated Jackson's theorem. 
\begin{theorem}[Jackson, \cite{davis1975interpolation}]\label{thm:Jackson}
For every $L$-lipschitz function $f:[-1,1]\to \mathbb{R}$ and $r\in\mathbb N$ there is a degree $r$ polynomial $p$  such that $\|p-f\|_{\infty,[-1,1]}\le \frac{6L}{r}$
\end{theorem}

\subsubsection{Localization}
An additional algorithmic component we will use, except polynomial regression, is {\em localization in the instance and the hypotheses space} (e.g. \citep{BartlettBoMe05,AwasthiBalcanLong14}). The basic idea is the following. Suppose that $w^*\in S^{d-1}$ defines the optimal halfspace. Suppose furthermore that we have found (say, using some simple algorithm) a vector $w\in S^{d-1}$ that defines a halfspace with a relatively small error. The facts that the marginal distribution is uniform and $\Err(h_w)$ is small have two relevant consequences:
\begin{itemize}
\item We know that the optimal vector, $w^*$, is close to $w$.
\item Hence, if $|\inner{w,x}|$ is large, then $h_{w^*}(x)=h_{w}(x)$ and therefore we know $h_{w^*}(x)$.
\end{itemize}
These two properties enable us to ``localize the learning" and concentrate only on hypotheses $h_{w'}$ with $w'$ close to $w$, and on instances $x$ with small $|\inner{w,x}|$. We will use this idea directly in our algorithm. In addition, we will use, as a black-box, the localization-based algorithm of \cite{AwasthiBalcanLong14}. Their algorithm starts with a crude approximation $w_1\in S^{d-1}$ of the optimal halfspace $w^{*}$. Then, it finds $w_2$ that minimizes the {\em hinge loss} $\mathrm{E}_{\cd|_{T\times \{\pm 1\}}}(1-\inner{w,yx})_+$ on the restriction of $\cd$ to some small strip $T=\{x\in S^{d-1}\mid |\inner{w,x}|\le \gamma\}$. Then, it continue in this manner to find better and better $w_i$'s. Awasthi, Balcan and Long used their algorithm to show:
\begin{theorem}\label{thm:ABL}\cite{AwasthiBalcanLong14}
There is an efficient learning algorithm with label complexity $\poly\left(d,\log\left(\frac{1}{\eta}\right)\right)$ that tolerates noise rate of $\frac{\eta}{\alpha_0}$ for some universal constant $\alpha_0>1$. Moreover, the algorithm is {\em proper}, that is, its output is a halfspace.
\end{theorem}

\subsubsection{The PTAS and its analysis}
In a nutshell, our algorithm first find (step \ref{alg_step_1}) a ``rough estimation", $w$, of $w^*$. Then, it ``localizes the learning" and apply more computation power (step \ref{alg_step_3}), to a small strip $T$ that is closed to $h_w$'s decision boundary, and therefore, intuitively, we are less certain about $h_w$'s prediction. 

\begin{algorithm}[H]
\caption{A PTAS for agnostically learning halfspaces w.r.t. the uniform distribution} \label{alg:PTAS}
\begin{algorithmic}
\STATE \textbf{Input:} $0<\eta\le 1$ and access to samples from a distribution $\cd$ on $S^{d-1}\times\{\pm 1\}$. 
\STATE \textbf{Parameters:} $r\in \mathbb N$, $\beta>0$ and $\gamma>0$.
\end{algorithmic}

\begin{algorithmic}[1]
\STATE\label{alg_step_1} Find, using \cite{AwasthiBalcanLong14} (theorem \ref{thm:ABL}), a vector $w\in S^{d-1}$ with $\Err_{\cd}(h_w)\le  \alpha_0\eta$
\STATE Let $T=T_{d,\gamma}(w):=\{u\in S^{d-1}: |\inner{w,u}|\le \gamma\}$.
\STATE\label{alg_step_3} Find, using \cite{KalaiKlMaSe05} (theorem \ref{thm:regression}), $P\in \mathbf{POL}_{r,d}$ with 
\[
\Err_{\cd|_T}(P)\le \Err_{\cd|_T}(h_{w^*})+\min_{P'\in \mathrm{POL}_{r,d}}\|h_{w^*}-P'\|_{1,\cd|_T}+\beta
\]
where $h_{w^*}$ is an optimal halfspace classifier w.r.t. $\cd$.
\STATE With probability $\frac{1}{2}$ return $h_w$, and w.p. $\frac{1}{2}$ return the classifier 
\[
h(x)=\begin{cases}
h_w(x) & |\inner{w,x}|>\gamma\\
\sign(P(x)) & |\inner{w,x}|\le \gamma
\end{cases}
\]
\end{algorithmic}
\end{algorithm}

\begin{theorem}[main -- detailed]\label{thm:main_detailed}
With appropriate choice of the parameters $r,\beta,\gamma$ (depending on $0<\mu,\eta\le 1$), algorithm \ref{alg:PTAS} satisfies:
\begin{itemize}
\item It tolerates noise rate of $(1-\mu)\eta$.
\item It runs in time $\poly\left(d^{\frac{\log^3\left(\frac{1}{\mu}\right)}{\mu^2}},\frac{1}{\eta}\right)$.
\item Its label complexity is $\poly\left(d^{\frac{\log^3\left(\frac{1}{\mu}\right)}{\mu^2}},\log\left(\frac{1}{\eta}\right)\right)$.
\end{itemize}
\end{theorem}
{\bf Proof outline.} To prove theorem \ref{thm:main_detailed}, we must show that we can choose the parameters so that the time and label complexity are as stated, and under the assumption that $\Err_{\cd}(h_{w^*})\le (1-\mu)\eta$, the error of the returned classifier satisfies $\Err_{\cd}(h)\le \eta$. Below, we explain how we do that.
We would naturally like to decompose the error into two parts:
\begin{eqnarray}\label{eq:exposition_1}
\Err_{\cd}(h) &=& \Pr_{(x,y)\sim \cd}\left(x\notin T\right)\cdot\Err_{\cd|_{T^c\times\{\pm 1\}}}(h)+\Pr_{(x,y)\sim \cd}\left(x\in T\right)\cdot\Err_{\cd|_{T\times\{\pm 1\}}}(h)
\nonumber \\
&=& \Pr_{(x,y)\sim \cd}\left(x\notin T\right)\cdot\Err_{\cd|_{T^c\times\{\pm 1\}}}(h_{w})+\Pr_{(x,y)\sim \cd}\left(x\in T\right)\cdot\Err_{\cd|_{T\times\{\pm 1\}}}(P)
\end{eqnarray}
We first handle the former summand using a localization lemma (lemma \ref{lem:localization} below). We show that for $\gamma=\Theta\left(\frac{\eta\sqrt{\log\left(\frac{1}{\mu}\right)}}{\sqrt{d}}\right)$, the probability that $h_w(x)\ne h_{w^*}(x)$ outside the strip $T$, is $\le \frac{\mu\eta}{2}$. Hence, on the complement of $T$, the returned classifier, that coincides with $h_w$, is as good as $h_{*}$, up to an additive error of $\frac{\mu\eta}{2}$. Concretely,
\begin{equation}\label{eq:exposition_2}
\Pr_{(x,y)\sim \cd}\left(x\notin T\right)\cdot\Err_{\cd|_{T^c\times\{\pm 1\}}}(h_{w})\le \Pr_{(x,y)\sim \cd}\left(x\notin T\right)\cdot\Err_{\cd|_{T^c\times\{\pm 1\}}}(h_{w^*})+\frac{\mu\eta}{2}~.
\end{equation}
It remains to handle the latter summand in equation (\ref{eq:exposition_2}). It is enough to show that 
\begin{equation}\label{eq:exposition_3}
\Pr_{(x,y)\sim T}\left(x\in T\right)\cdot\Err_{\cd|_{T\times\{\pm 1\}}}(P)\le \Pr_{(x,y)\sim T}\left(x\in T\right)\cdot\Err_{\cd|_{T\times\{\pm 1\}}}(h_{w^*})+\frac{\mu\eta}{2}
\end{equation}
Indeed, in that case it follows from equations (\ref{eq:exposition_1}), (\ref{eq:exposition_2}) and (\ref{eq:exposition_3}) that
\begin{eqnarray*}
\Err_{\cd}(h) &\le& \Pr_{(x,y)\sim \cd}\left(x\notin T\right)\cdot\Err_{\cd|_{T^c\times\{\pm 1\}}}(h_{w^*})+\Pr_{(x,y)\sim T}\left(x\in T\right)\cdot\Err_{\cd|_{T\times\{\pm 1\}}}(h_{w^*})+\mu\eta
\\
&=& \Err_{\cd}(h_{w^*})+\mu\eta\le (1-\mu)\eta+\mu\eta=\eta~.
\end{eqnarray*}
To prove equation (\ref{eq:exposition_3}) we first note that $\Pr_{(x,y)\sim \cd}(x\in T)=\Theta\left(\eta\sqrt{\log\left(\frac{1}{\mu}\right)}\right)$. Hence, it is enough to show that for suitable choice of $r$ and $\beta$, $\Err_{\cd|_{T\times\{\pm 1\}}}(P)\le \Err_{\cd|_{T\times\{\pm 1\}}}(h_{w^*})+\frac{\mu}{C \sqrt{\log\left(\frac{1}{\mu}\right)}}$ for large enough constant $C>0$. 
By theorem \ref{thm:regression}, it is enough to choose $\beta=\frac{\mu}{2C \sqrt{\log\left(\frac{1}{\mu}\right)}}$, and large enough $r$ so that $\min_{P'\in \POL_{r,d}}\|h-P'\|_{1,\cd|_{T\times\{\pm 1\}}}\le \frac{\mu}{2C \sqrt{\log\left(\frac{1}{\mu}\right)}}$.

As we show, $r=O\left(\frac{\log^{3}\left(\frac{1}{\mu}\right)}{\mu^2}\right)$ suffices. To do that, by equation (\ref{eq:appr_half_by_appr_sign}), it is enough to find a polynomial of degree $O\left(\frac{\log^{3}\left(\frac{1}{\mu}\right)}{\mu^2}\right)$ that approximates the sign function up to an $\ell_1$-error of $\frac{\mu}{2C\sqrt{\log\left(\frac{1}{\mu}\right)}}$ w.r.t the distribution $(\cd|_{T\times\{\pm 1\}})_{w^*}$.
This is done in section \ref{sec:sign_appr}, in three steps:
\begin{enumerate}
\item We first (section \ref{sec:sign_appr_1}) show how to find polynomials that approximate the sign function on all the points of a given segment $[-a,a]$, except the area that is very close to the origin, say $[-\epsilon,\epsilon]$. To this end, we invoke Jackson's theorem (theorem \ref{thm:Jackson}) to find a polynomial that roughly (up to an error of, say, $0.1$) approximates the sign function on the mentioned regime. Namely, we find a polynomial $p$ of degree $O\left(\frac{a}{\epsilon}\right)$ that maps $[-a,-\epsilon]$ (resp. $[\epsilon,a]$) to $[-1.1,-0.9]$ (resp. $[0,9,1.1]$). To move from accuracy of $0.1$ to accuracy of some small $\tau>0$, we compose $p$ with another polynomial $r$ that maps $[-1.1,-0.9]$ (resp. $[0.9,1.1]$) to $[-1-\tau,-1+\tau]$ (resp. $[1-\tau,1+\tau]$). Using the Taylor expansion of the the {\em error function} \mbox{$\erf(x):=\frac{1}{\sqrt{2\pi}}\int_{-\infty}^xe^{-\frac{t^2}{2}}dt$}, we show that there exists such $r$ of degree $O\left(\log\left(\frac{1}{\tau}\right)\right)$.
\item In the second step (section \ref{sec:sign_appr_2}), we will find $\ell_1$ approximations for distributions with strong tail bounds (namely, distributions that have density function bounded by $2\exp\left(-\frac{x^2}{32}\right)$ on a certain domain). We use step 1 to find polynomials that approximate the sign function in $\ell_\infty$ on a relatively large area, and use the tail bounds and lemma \ref{lem:val_bound} to neglect the $\ell_1$ norm on the complement of that area.
\item In the last step (section \ref{sec:sign_appr_3}), using basic facts about high dimensional spherical geometry, we show that the distribution $(\cd|_{T\times\{\pm 1\}})_{w^*}$ have strong enough tail bounds.
\end{enumerate}

\subsection{Related work}\label{sec:related_work}
{\bf Upper bounds.} Statistical aspects of learning halfspaces have been extensively studied (e.g. \cite{Vapnik98}).
Halfspaces are efficiently learnable in the {\em realizable case}, when $\opt=0$. This is done using the ERM algorithm \cite{Vapnik98} that efficiently find, using linear programming, a halfspace that makes no errors on the given sample.
For agnostic, distribution free learning, the best known efficient algorithm \cite{KearnsLi88} have an approximation ratio of $O(d)$, and the best known exact 
algorithm is the naive (exponential time) algorithm that go over all halfspaces and return the one with minimal error on the given sample. Under distributional 
assumptions, better algorithms are known. Under the uniform distribution, \cite{KalaiKlMaSe05} and \cite{AwasthiBalcanLong14} presented efficient algorithms 
with approximation ratios $\sqrt{\log\left(\frac{1}{\opt}\right)}$ and $O(1)$ respectively. The best known exact 
algorithm \cite{KalaiKlMaSe05} runs in time $d^{O\left(\frac{1}{\epsilon^2}\right)}$ (as follows from \cite{diakonikolas2010bounded2}). 
For log-concave distributions, \cite{KlivansLoSe09} and \cite{AwasthiBalcanLong14} presented efficient algorithms 
with approximation ratios $O\left(\frac{\log\left(\frac{1}{\opt}\right)}{\opt^2}\right)$
and $O\left(\log^2\left(\frac{1}{\opt}\right)\right)$ respectively. The best known exact 
algorithm \cite{KalaiKlMaSe05} runs in time $d^{f(\epsilon)}$.
In learning halfspaces with margin\footnote{In this problem the distribution is supported in the unit ball, and the algorithm should compete with all classifiers that predict like a halfspace classifier $h_{w}$, except that they give no prediction (and therefore err) for instances that are within distance $\gamma$ of the decision boundary of $h_w$.} $\gamma>0$, best known algorithms \cite{LongSe11,BirnbaumSh12} have approximation ratio of $\frac{1/\gamma}{\log\left(1/\gamma\right)}$, while the best known exact algorithm \cite{ShalevShSr11} runs in time $\left(\frac{1}{\epsilon}\right)^{O\left(\frac{\log\left(1/\gamma\right)}{\gamma}\right)}$.
\\
{\bf Lower bounds.} Hardness of (distribution free)  agnostic learning of halfspaces is known to follow from several complexity assumptions including hardness of learning parity \cite{KalaiKlMaSe05} (this result even rules out learning under the uniform distribution on $\{\pm 1\}^d$), hardness of the shortest vector problem \cite{FeldmanGoKhPo06}, and hardness of refuting random $K$-SAT formulas \cite{danielyImp3}. Hardness of learning sparse parity implies hardness of agnostic learning under the uniform distribution on $S^{d-1}$ \cite{klivans2014embedding}.
For every $\tau>0$, hardness of agnostic learning of halfspaces with an approximation ratio of $2^{\log^{1-\tau}(d)}$ follows from 
hardness of refuting random $K$-XOR formulas \cite{danielyImp4} (see also \cite{danielyImp2}).
For {\em proper} learning of halfspaces, super constant ($2^{\log^{1-\tau}(d)}$ for every $\tau>0$) lower bounds on the best approximation ratio are known, assuming $\mathcal{NP}\ne\mathcal{RP}$ \cite{arora1993hardness, GuruswamiRa06, FeldmanGoKhPo06}. Finally, lower bounds on concrete families of algorithms were studied in \cite{Ben-DavidLoSrSr12,danielyLin}

\section{Proof of theorem \ref{thm:main_detailed}}\label{sec:proof}
For localization arguments, we will use the following lemma.
\begin{lemma}[localization]\label{lem:localization}
Let $w,w^*\in S^{n-1}$ and let $\cd$ be a distribution of $S^{d-1}\times \{\pm 1\}$ such that $\cd|_{S^{d-1}}$ is uniform.
\begin{itemize}
\item We have $\frac{\theta(w,w^*)}{\pi}\le \Err_{\cd}(w)+\Err_{\cd}(w^*)$.
\item If $x\in S^{d-1}$ is a uniform vector, then for every $r>0$,
\[
\Pr\left(h_w(x)\ne h_{w^*}(x)\text{ and }|\inner{x,w}|>r\cdot \theta(w,w^*)\right)\le \frac{4\cdot\theta(w,w^*)}{\pi}\exp\left(-\frac{1}{8}r^2d\right)
\]
\end{itemize}
\end{lemma}
\proof
For the first part we note that $\Pr_{x\sim\cd}\left(h_w(x)\ne h_{w^*}(x)\right)=\frac{\theta(w,w^*)}{\pi}$, while on the other hand,
\[
\Pr_{x\sim\cd}\left(h_w(x)\ne h_{w^*}(x)\right)\le \Pr_{(x,y)\sim\cd}\left(h_w(x)\ne y\right)+\Pr_{(x,y)\sim\cd}\left(h_{w^*}(x)\ne y\right)~.
\]
For the second part, let $V\subset\mathbb{R}^d$ be the $2$-dimensional space spanned by $w,w^*$, let $P_V:\mathbb{R}^d\to V$ be the orthogonal projection $V$, and let $B\subset V$ be the ball of radius $r$ around $0$. We have
\[
|\inner{w^*,x}- \inner{w,x}| = |\inner{w^*-w,P_V(x)}|\le \|w-w^*\|\cdot \|P_V(x)\|\le \theta(w,w^*)\cdot \|P_V(x)\|~.
\]
Therefore, if $P_V(x)\in B$ and $|\inner{x,w}|> r\cdot\theta(w,w^*)$ then $h_w(x)=h_{w^*}(x)$. It follows that
\begin{eqnarray*}
\Pr\left(h_w(x)\ne h_{w^*}(x)\text{ and }|\inner{x,w}|>r\cdot\theta(w,w^*)\right)
&=&
\Pr\left(h_w(x)\ne h_{w^*}(x)\mid P_V(x)\notin B \right)\cdot\Pr\left(P_V(x)\notin B\right)
\\
&=&
\frac{\theta(w,w^*)}{\pi}\cdot\Pr\left(P_V(x)\notin B\right)~.
\end{eqnarray*}
Finally, let $e_1,e_2\in V$ be an orthonormal basis. Note that if $|\inner{x,e_1}|\le \frac{r}{\sqrt{2}}$ and $|\inner{x,e_2}|\le \frac{r}{\sqrt{2}}$ then $P_V(x)\in B$. Hence, we have
\[
\Pr\left(P_V(x)\notin B\right)\le \Pr\left( |\inner{x,e_1}|> \frac{r}{\sqrt{2}} \right)+
\Pr\left( |\inner{x,e_2}|> \frac{r}{\sqrt{2}} \right)\le 4\exp\left(-\frac{1}{8}r^2d\right)~.
\]
Here, the last inequality follows from the well known measure concentration bound according which for every $e\in S^{d-1}$ and $\sigma>0$ we have $\Pr\left(|\inner{x,e}|\ge \sigma \right)\le 2\exp\left(-\frac{1}{4}\sigma^2d\right)$.
\proofbox
To approximate $h_{w^*}$, we will need to find low degree $\ell_1$ approximation of $h_{w^*}$ w.r.t. the distribution $\cd|_T$. Such approximations are given in the following two lemmas.
The first lemma is from \cite{diakonikolas2010bounded} (see a proof in section \ref{sec:sign_appr}. For a stronger version, with $r=O\left(\frac{1}{\tau^2}\right)$, see \cite{diakonikolas2010bounded2}).
The proof of the second lemma is established by approximating the sign function (as explained in section \ref{sec:sign_appr_intro}) and is given in section \ref{sec:sign_appr}.
\begin{lemma}[uniform halfspaces approximation, \cite{diakonikolas2010bounded}]\label{lem:halfspace_appr_no_restriction} Let $\cd$ be the uniform distribution on $S^{d-1}$ and let $w^{*}\in S^{d-1}$. For every $\tau>0$ there is $P\in\POL_{r,d}$, for $r=O\left(\frac{\log^2\left(1/\tau\right)}{\tau^2}\right)$ such that $\|h_{w^*}-P\|_{1,\cd}<\tau$.
\end{lemma}

\begin{lemma}[halfspaces approximation on a strip]\label{lem:halfspace_appr_strip} Let $w,w^*$ be two vectors with $\theta=\theta(w,w^*)$ and let $\frac{1}{2}>\gamma>0$. Let $\cd$ be the distribution on $S^{d-1}$ that is the restriction of the uniform distribution to $T_{d,\gamma}(w)$. Then, for every $0<\tau<\frac{\sin(\theta)}{2\gamma\sqrt{d}}$ there is $P\in\POL_{r,d}$, for $r=O\left(\frac{\log^2\left(1/\tau\right)}{\tau^2}\right)$ such that $\|h_{w^*}-P\|_{1,\cd}<\tau$.
\end{lemma}
Lastly, we will also rely on the following complexity analysis of algorithm \ref{alg:PTAS}.

\begin{lemma}[complexity analysis]\label{lem:PTAS_complexity}
The runtime of algorithm \ref{alg:PTAS} is $\poly\left(d^{r},\frac{1}{\beta},\frac{1}{\gamma},\frac{1}{\eta}\right)$ and the label complexity is $\poly\left(d^{r},\frac{1}{\eta},\log\left(\frac{1}{\eta}\right)\right)$.
\end{lemma}
\proof The runtime of step \ref{alg_step_1} is $\poly\left(d,\frac{1}{\eta}\right)$, while the label complexity is $\poly\left(d,\log\left(\frac{1}{\eta}\right)\right)$. For step \ref{alg_step_3}, we can apply the \cite{KalaiKlMaSe05} algorithm on $\poly\left(d^{r},\frac{1}{\eta}\right)$ examples and labels from the distribution $\cd|_T$. We can get these many examples by sampling $\poly\left(d^{r},\frac{1}{\beta},\frac{1}{\Pr_{\cd}(T\times \{\pm 1\})}\right)$ examples from $\cd$ and keep and expose the labels of only the first $\poly\left(d^{r},\frac{1}{\beta}\right)$ examples that fell in $T$. It is not hard to see that $\Pr_{\cd}(T\times \{\pm 1\})\ge \Omega\left(\min\left(\gamma\sqrt{d},1\right)\right)$. Hence, the runtime of step \ref{alg_step_3} is $\poly\left(d^{r},\frac{1}{\beta},\frac{1}{\gamma}\right)$. To summarize, the total runtime is $\poly\left(d^{r},\frac{1}{\beta},\frac{1}{\gamma},\frac{1}{\eta}\right)$ and the label complexity is $\poly\left(d^{r},\frac{1}{\beta},\log\left(\frac{1}{\eta}\right)\right)$.
\proofbox
We are now ready to prove theorem \ref{thm:main_detailed}.

\proof (of theorem \ref{thm:main_detailed})
We will first deal with the case that $\eta>\frac{1}{2(1+\alpha_0)}$. In that case we won't use localization, that is we will choose $\gamma=1$ (in that case our algorithm is essentially the algorithm of \cite{KalaiKlMaSe05}). We will choose $\beta=\frac{\mu\eta}{2}$, and $r=O\left(\frac{\log^2\left(1/(\mu\eta)\right)}{(\mu\eta)^2}\right)=O\left(\frac{\log^2\left(1/\mu\right)}{\mu^2}\right)$ that is large enough so that $\min_{P'\in \mathrm{POL}_{r,d}}\|h_{w^*}-P'\|_{1,\cd|_T}\le \frac{\mu\eta}{2}$ (this is possible according to lemma \ref{lem:halfspace_appr_no_restriction}). It that case, the algorithm will, w.p. $\frac{1}{2}$, return the hypothesis $\sign(P)$ for the polynomial $P$ that was found in step \ref{alg_step_3}. We have
\[
\Err_{\cd}(P)\le \Err_{\cd}(h_{w^*})+\frac{\mu\eta}{2}+\frac{\mu\eta}{2}~.
\]
By assumption, $\Err_{\cd}(h_{w^*})\le (1-\mu)\eta$. Hence, $\Err_{\cd}(P)\le \eta$, as required. It also follows from lemma \ref{lem:PTAS_complexity} that the runtime and label complexity are $\poly\left(d^{\frac{\log^2\left(1/\mu\right)}{\mu^2}}\right)$ (note that $\eta$ is bounded from below by a constant) as stated .

Next, we deal with the case that $\eta\le\frac{1}{2(1+\alpha_0)}$. We will show that it is possible to choose $r=\Theta\left(\frac{\log^3\left(\frac{1}{\mu}\right)}{\mu^2}\right)$, $\beta=\theta\left(\frac{\mu}{\sqrt{\log\left(\frac{1}{\mu}\right)}}\right)$ and $\gamma=\Theta\left(\frac{\eta\sqrt{\log\left(\frac{1}{\mu}\right)}}{\sqrt{d}}\right)$ for which the algorithm will have the desired properties. Also, by lemma \ref{lem:PTAS_complexity}, for such a choice of parameters, the runtime and label complexity are as stated.

Let $w^*$ be the vector defining the optimal halfspace. By assumption, $\Err_{\cd}(h_{w^*})\le (1-\mu)\eta$. Let $w$ be the vector found in step \ref{alg_step_1}, and let $P$ be the polynomial found in step \ref{alg_step_3}. We first claim that we can assume w.l.o.g. that
\begin{equation}\label{eq:8}
\frac{\theta}{\pi}:=\frac{\theta(w,w^*)}{\pi}\ge \mu\eta~.
\end{equation}
Indeed, otherwise, we will have
\begin{eqnarray*}
\Err_{\cd}(h_w)&\le& \Err_{\cd}(h_{w^*})+\Pr_{(x,y)\sim\cd}\left(h_w(x)\ne h_{w^*}(x)\right)
\\
&=& \Err_{\cd}(h_{w^*})+\frac{\theta}{\pi}\le (1-\mu)\eta+\mu\eta<\eta
\end{eqnarray*}
and in that case the algorithm will return, in the last step, w.p. $\frac{1}{2}$, a hypothesis with error at most $\eta$, as required.

Let \mbox{$h(x)=\begin{cases} h_w(x) & |\inner{w,x}|>\gamma\\ \sign(P(x)) & |\inner{w,x}|\le \gamma\end{cases}$}. It is enough to show that $\Err_{\cd}(h)\le\eta$.
Let $T=T_{d,\gamma}(w):=\{u\in S^{d-1}: |\inner{w,u}|\le \gamma\}$.
The error of $h$ is
\begin{eqnarray}\label{eq:3}
\Err_{\cd}(h) &=& 
\Pr_{(x,y)\sim \cd}\left(h_w(x)\ne y\text{ and }|\inner{w,x}|>\gamma\right)+\Pr_{(x,y)\sim\cd}\left(\sign(P(x))\ne y\text{ and }|\inner{w,x}|\le\gamma\right)
\nonumber \\
&\le &\Pr_{(x,y)\sim \cd}\left(h_w(x)\ne h_{w^*}(x)\text{ and }|\inner{w,x}|>\gamma\right)+\Pr_{(x,y)\sim \cd}\left(h_{w^*}(x)\ne y\text{ and }|\inner{w,x}|>\gamma\right)
\nonumber \\
&&+\Pr_{(x,y)\sim\cd}\left(x\in T\right)\cdot \Err_{\cd|_T}(P)
\end{eqnarray}
By the first part of lemma \ref{lem:localization} we have
\begin{equation}\label{eq:7}
\frac{\theta}{\pi}\le \Err_{\cd}(h_w)+\Err_{\cd}(h_{w^*})\le (1+\alpha_0)\eta~.
\end{equation}
By the second part of lemma \ref{lem:localization} we have
\begin{eqnarray*}
\Pr_{(x,y)\sim \cd}\left(h_w(x)\ne h_{w^*}(x)\text{ and }|\inner{w,x}|>\gamma\right)&\le & 4(1+\alpha_0)\eta\exp\left(-\frac{1}{8} \left(\frac{\gamma}{\theta}\right)^2 d\right)
\\
&\le & 4(1+\alpha_0)\eta\exp\left(-\frac{1}{8} \left(\frac{\gamma}{(1+\alpha_0)\pi\eta}\right)^2 d\right)
\end{eqnarray*}
Now, by an appropriate choice of $\gamma=\Theta\left(\frac{\eta\sqrt{\log\left(\frac{1}{\mu}\right)}}{\sqrt{d}}\right)$, we get
\begin{equation}\label{eq:4}
\Pr_{(x,y)\sim \cd}\left(h_w(x)\ne h_{w^*}(x)\text{ and }|\inner{w,x}|>\gamma\right)\le\frac{\mu\eta}{2}~.
\end{equation}
We next deal with the term $\Pr_{(x,y)\sim\cd}\left(x\in T\right)\cdot \Err_{\cd|_T}(P)$. Since $\gamma=\Theta\left(\frac{\eta\sqrt{\log\left(\frac{1}{\mu}\right)}}{\sqrt{d}}\right)$ we have that 
\begin{equation}\label{eq:6}
\Pr_{(x,y)\sim\cd}\left(x\in T\right)=O\left( \eta\cdot\sqrt{\log\left(\frac{1}{\mu}\right)}\right)
\end{equation}
Also, by equation (\ref{eq:7}) and the assumption that $\eta\le\frac{1}{2(\alpha_0+1)}$, we have that $0\le \theta\le \frac{\pi}{2}$. For this regime, $\sin(\theta)\ge \frac{2\theta}{\pi}$. Hence, by equation (\ref{eq:8}) we have
\begin{equation}\label{eq:9}
\frac{\sin(\theta)}{2\gamma\sqrt{d}}\ge \frac{\theta}{\pi\gamma\sqrt{d}}\ge \frac{\mu\eta}{\gamma\sqrt{d}}=\Theta\left(\frac{\mu}{\sqrt{\log\left(1/\mu\right)}}\right)
\end{equation}
By equations (\ref{eq:6}) and (\ref{eq:9}) we can choose $\beta=\frac{\mu}{4C\sqrt{\log\left(\frac{1}{\mu}\right)}}$, where $C>0$ is a universal constant that is large enough so that 
\begin{equation}\label{eq:10}
\beta<\frac{\sin(\theta)}{2\gamma\sqrt{d}}\text{ and }2\beta\cdot \Pr_{(x,y)\sim\cd}\left(x\in T\right)\le \frac{\mu\eta}{2}
\end{equation}
By equation \ref{eq:10} and lemma \ref{lem:halfspace_appr_strip} we can choose $r=\Theta\left(\frac{\log^2\left(\frac{1}{\beta}\right)}{\beta^2}\right)=\Theta\left(\frac{\log^3\left(\frac{1}{\mu}\right)}{\mu^2}\right)$ such that
\[
\min_{P'\in \mathrm{POL}_{r,d}}\|h_{w^*}-P'\|_{1,\cd|_T}\le \beta
\]
in that case we have 
\[
\Err_{\cd|_T}(P)\le \Err_{\cd|_T}(h_{w^*})+\min_{P'\in \mathrm{POL}_{r,d}}\|h_{w^*}-P'\|_{1,\cd|_T}+\beta\le
\Err_{\cd|_T}(h_{w^*})+2\beta ~.
\]
Hence,
\begin{eqnarray}\label{eq:5}
\Pr_{(x,y)\sim\cd}\left(x\in T\right)\cdot \Err_{\cd|_T}(P) 
&\le & \Pr_{(x,y)\sim\cd}\left(x\in T\right)\cdot\Err_{\cd|_T}(h_{w^*}) + \Pr_{(x,y)\sim\cd}\left(x\in T\right)\cdot 2\beta
\nonumber \\
&\le& \Pr_{(x,y)\sim\cd}\left(x\in T\right)\cdot\Err_{\cd|_T}(h_{w^*}) + \frac{\mu\eta}{2}
\nonumber \\
&=&  \Pr_{(x,y)\sim \cd}\left(h_{w^*}(x)\ne y\text{ and }|\inner{w,x}|\le\gamma\right)+ \frac{\mu\eta}{2}
\end{eqnarray}
By equations (\ref{eq:3}), (\ref{eq:4}) and (\ref{eq:5}) we conclude that
\begin{eqnarray*}
\Err_{\cd}(h) &\le& 
\frac{\mu\eta}{2}+\Pr_{(x,y)\sim \cd}\left(h_{w^*}(x)\ne y\text{ and }|\inner{w,x}|>\gamma\right)
\nonumber \\
&&+\Pr_{(x,y)\sim \cd}\left(h_{w^*}(x)\ne y\text{ and }|\inner{w,x}|\le\gamma\right)+ \frac{\mu\eta}{2}
\nonumber \\
&=& \Err_{\cd}(h_{w^*})+\mu\eta\le (1-\mu)\eta+\mu\eta=\eta~.
\end{eqnarray*}
\proofbox

\section{Polynomial approximation of the sign function}\label{sec:sign_appr}
In this section we will find $\ell_1$ approximation of halfspaces. In particular, we will prove lemmas \ref{lem:halfspace_appr_strip} and \ref{lem:halfspace_appr_no_restriction}. 

\subsection{Approximation in ``truncated $L^\infty$"}\label{sec:sign_appr_1}
\begin{lemma}\label{lem:sig_appr_l_inf}
Let $a,\gamma,\tau>0$. There exist a polynomial $p$ of degree $O\left(\frac{1}{\gamma}\cdot \log\left(\frac{1}{\tau}\right)\right)$ such that 
\begin{itemize}
\item For $x\in [-a,a]$, $|p(x)|<1+\tau$.
\item For $x\in [-a, a]\setminus [-\gamma\cdot a, \gamma\cdot a]$, $|p(x)-\sign(x)|<\tau$.
\end{itemize}
\end{lemma}
We will use the following lemma:
\begin{lemma}\label{lem:sig_appr_l_inf_sublemma}
Let $\tau>0$. There exist a polynomial $p$ of degree $O\left(\log\left(\frac{1}{\tau}\right)\right)$ such that 
\begin{itemize}
\item For $x\in [-1.5,1.5]$, $|p(x)|<1+\tau$.
\item For $x\in [-1.5,1.5]\setminus [-0.5,0.5]$, $|p(x)-\sign(x)|<\tau$.
\end{itemize}
\end{lemma}
\proof
The proof is established by approximating the error function, $\erf(x):=\frac{1}{\sqrt{2\pi}}\int_{-\infty}^xe^{-\frac{t^2}{2}}dt$ by a low degree polynomial.
Let $\sigma=2\sqrt{2\log(\frac{4}{\sqrt{2\pi}\tau})}$. We claim that for every $x>\frac{\sigma}{2}$ we have
\begin{equation}\label{eq:1}
|\erf(x)- 1|,|\erf(-x)|\le \frac{\tau}{4}~.
\end{equation}
Because $0\le \erf(x)\le 1$ for all $x$, and since $\erf(x)=1-\erf(-x)$, it is enough to prove that 
$\erf(x)\ge 1-\frac{\tau}{4}$. Indeed, we have
\begin{eqnarray*}
1-\erf(x) &=& \frac{1}{\sqrt{2\pi}}\int_{x}^\infty e^{-\frac{t^2}{2}}dt\\
&\le &\frac{1}{\sqrt{2\pi}}\int_{x}^\infty te^{-\frac{t^2}{2}}dt\\
&=& \frac{1}{\sqrt{2\pi}}\left[ \left.  -e^{-\frac{t^2}{2}}\right\vert_{x}^\infty\right]\\
&= & \frac{1}{\sqrt{2\pi}}e^{-\frac{x^2}{2}}
\\
&\le & \frac{1}{\sqrt{2\pi}}e^{-\frac{\sigma^2}{2}}=\frac{\tau}{4}~.
\end{eqnarray*}
Now, by the Taylor expansion of $e^x$ we have
\[
e^{-\frac{x^2}{2}}=\sum_{n=0}^\infty \frac{(-1)^nx^{2n}}{n!2^n}~.
\]
Integrating element-wise and using the fact that $\erf(0)=\frac{1}{2}$, we have
\[
\erf(x)=\frac{1}{\sqrt{2\pi}}\int_{-\infty}^xe^{-\frac{t^2}{2}}dt=\frac{1}{2}+\frac{1}{\sqrt{2\pi}}\sum_{n=0}^\infty \frac{(-1)^nx^{2n+1}}{n!2^n(2n+1)}~.
\]
Let $r$ be the $2k$'th Taylor polynomial of $\erf$ for $k=\max\{\lceil 2(1.5\sigma)^2e\rceil,\log_2\left(\frac{4}{\tau}\right)\}=O\left(\log\left(\frac{1}{\tau}\right)\right)$. We have, for $|x|\le 1.5\sigma\le \sqrt{\frac{k}{2e}}$
\begin{eqnarray*}
|r(x)-\erf (x)| &\le & \frac{2}{\sqrt{\pi}}\sum_{n=k}^\infty\frac{|x|^{2n+1}}{n!(2n+1)}\\
&\le & \frac{2}{\sqrt{\pi}}\sum_{n=k}^\infty\frac{x^{2n}}{n!}\\
&\le & \frac{2}{\sqrt{\pi}}\sum_{n=k}^\infty\frac{x^{2n}}{\sqrt{2\pi}\left(\frac{n}{e}\right)^n}\\
&\le & \frac{\sqrt{2}}{\pi}\sum_{n=k}^\infty \left(\frac{x^2e}{n}\right)^n\\
&\le & \frac{\sqrt{2}}{\pi}\sum_{n=k}^\infty \left(\frac{1}{2}\right)^n\\
&= & \frac{\sqrt{2}}{\pi} \left(\frac{1}{2}\right)^{k-1} \le \left(\frac{1}{2}\right)^{k}\le \frac{\tau}{4}
\end{eqnarray*}
Here, the 4'th inequality follows from the well known fact that $n!\ge \sqrt{2\pi} \left(\frac{n}{e}\right)^{n}$.
Finally, using the last inequality and equation 
(\ref{eq:1}), it is not hard to check that the polynomial $p(x)=2r(\sigma x)-1$ satisfies the required properties.
\proofbox

\proof (of lemma \ref{lem:sig_appr_l_inf})
By rescaling, we can assume w.l.o.g. that $a=1$.
Let $\phi:[-1,1]:\to \reals$ be the function
$$\phi(x)=\begin{cases}\frac{1}{\gamma}x & |x|\le\gamma \\ 1 & x\ge \gamma \\ -1 & x\le -\gamma\end{cases}$$
By Jackson's Theorem, there is a polynomial $q:[-1,1]\to \reals$ of degree  $\le \left\lceil \frac{12}{\gamma} \right\rceil$ with \mbox{$||q-\phi||_{\infty,[-1,1]}\le \frac{1}{2}$}. Also, let $r$ be the polynomial from Lemma \ref{lem:sig_appr_l_inf_sublemma}.
It is easy to check that, $p=r\circ q$ satisfies the requirement of the Lemma.
\proofbox

\subsection{Approximations for short tailed distributions}\label{sec:sign_appr_2}
\begin{lemma}\label{lem:sig_appr_sub_gauss}
Let $\rho:\mathbb R\to \mathbb R_+$ a density function such that for some $\gamma,\sigma>0$ we have
\[
\forall x,\;\rho(x)\le \frac{2}{\sigma}\text{ and } \forall |x|>2\gamma,\;\rho(x)\le \frac{2}{\sigma}\exp\left(-\frac{x^2}{32\sigma^2}\right)
\]
Then, for every $0<\tau\le \frac{\sigma}{2\gamma}$ there is a polynomial of degree\footnote{The constant in the big-O notation is universal.} $O\left(\frac{\log^2\left(1/\tau\right)}{\tau^2}\right)$ such that
\[
\int_{-\infty}^{\infty} |p(x)-\sign(x)|\rho(x)dx\le \tau
\]
\end{lemma}
We will use the following fact.
\begin{lemma}\cite{beneliezer2009}\label{lem:val_bound}
Let $p:\mathbb R\to\mathbb R$ be a polynomial of degree $\le r$ for which $|p(x)|\le b$ in the interval $[-a,a]$. Then, for every $|x|\ge a$ we have $|p(x)|\le b\cdot \left|\frac{2x}{a}\right|^r$.
\end{lemma}

\proof (of lemma \ref{lem:sig_appr_sub_gauss})
By lemma \ref{lem:sig_appr_l_inf}, there is a polynomial $p$ of degree $O\left(r\log\left(1/\tau\right)\right)$ such that
\begin{itemize}
\item For $x\in \left[-r\tau\sigma,r\tau\sigma\right]$, $|p(x)|<2$.
\item For $x\in \left[-r\tau\sigma,-\frac{\tau\sigma}{100}\right]$, $|p(x)|<\frac{\tau}{100}$.
\item For $x\in \left[\frac{\tau\sigma}{100},r\tau\sigma\right]$, $|p(x)-1|<\frac{\tau}{100}$.
\end{itemize}
We have, 
\begin{eqnarray*}
\int_{-\infty}^{\infty} |p(x)-\sign(x)|\rho(x)dx &=& 
\int_{|x|<\frac{\tau\sigma}{100}} |p(x)-\sign(x)|\rho(x)dx+
\int_{\frac{\tau\sigma}{100} \le |x| \le r\tau\sigma} |p(x)-\sign(x)|\rho(x)dx
\\
&&+\int_{|x|\ge r\tau\sigma}  |p(x)-\sign(x)|\rho(x)dx
\\
&\le &\int_{|x|<\frac{\tau\sigma}{100}} \frac{6}{\sigma}dx+
\int_{\frac{\tau\sigma}{100} \le |x| \le r\tau\sigma} \frac{\tau}{100}\rho(x)dx
\\
&&+\int_{|x|\ge r\tau\sigma}  |p(x)-\sign(x)|\rho(x)dx
\\
&\le & \frac{\tau}{2}+\int_{|x|\ge r\tau\sigma}  |p(x)-\sign(x)|\rho(x)dx
\end{eqnarray*}
It remains to bound $\int_{|x|\ge r\tau\sigma}  |p(x)-\sign(x)|\rho(x)dx$. 
We will choose $r\ge\frac{1}{\tau^2}$, and therefore we will have $r\tau\sigma\ge \frac{\sigma}{\tau}\ge 2\gamma$. Hence, by lemma \ref{lem:val_bound} we have
\begin{eqnarray*}
\int_{|x| \ge r\tau\sigma}  |p(x)-\sign(x)|\rho(x)dx
&\le &
\int_{|x| \ge r\tau\sigma} 3\left(\frac{2x}{r\tau\sigma}\right)^r\frac{2}{\sigma} e^{-\frac{x^2}{32\sigma^2}}dx
\\
&\le &
12\int_{r\tau\sigma}^{\infty} \left(\frac{2x}{r\tau\sigma}\right)^r\frac{1}{\sigma} e^{-\frac{x^2}{32\sigma^2}}dx
\\
&= &
12\int_{r\tau}^{\infty} \left(\frac{2y}{r\tau}\right)^re^{-\frac{y^2}{32}}dy 
\\
&\le &
12\int_{r\tau}^{\infty} \left(\left(\frac{2y}{r\tau}\right)^re^{-\frac{y^2}{64}}\right)e^{-\frac{y^2}{64}}dy 
\end{eqnarray*}
Now, it is possible to choose $r=\Theta\left(\frac{\log\left(1/\tau\right)}{\tau^2}\right)$ such that for all $y>r\tau$ we have $\left(\frac{2y}{r\tau}\right)^r\cdot e^{-\frac{y^2}{64}}\le 1$. For such $r$, the last expression is bounded by $12\int_{\omega\left(\frac{1}{\tau}\right)}^{\infty} e^{-\frac{y^2}{64}}dy=o(\tau)$.
\proofbox

\subsection{Approximation on a biased strip: proof of lemma \ref{lem:halfspace_appr_strip}}\label{sec:sign_appr_3}
In this section we will find a low degree approximation of halfspaces w.r.t. to the distribution from step \ref{alg_step_3} of our PTAS. Namely, we will prove lemma \ref{lem:halfspace_appr_strip}. Let $\rho_{d,\gamma,\theta}:[-1,1]\to \mathbb R_+$ be the projection on $w^*$ of the uniform distribution on $T_{d,\gamma}(w)$. By equation (\ref{eq:appr_half_by_appr_sign}), it is enough to find $\tau$-approximation of the sign function in $\ell_1$, w.r.t. $\rho_{d,\gamma,\theta}$. Namely, it is enough to prove:
\begin{lemma}\label{lem:halfspace_appr_strip_restated}
There is a univariate polynomial $p$ of degree $r=O\left(\frac{\log^2\left(1/\tau\right)}{\tau}\right)$ such that
\[
\int_{-1}^1|\sign(x)-p(x)|\rho_{d,\gamma,\theta}(x)dx\le \tau~.
\]
\end{lemma}
Lemma \ref{lem:halfspace_appr_strip_restated} follows immediately from lemma \ref{lem:sig_appr_sub_gauss} with $\sigma=\frac{\sin(\theta)}{\sqrt{d}}$, the assumptions that $\gamma<\frac{1}{2}$ and $\tau<\frac{\sin(\theta)}{2\gamma\sqrt{d}}$, and the following bound:

\begin{lemma}\label{lem:bound_on_truncated_dist}
\[
\forall z,\; \rho_{d,\gamma,\theta}(z)\le \frac{\sqrt{d}}{\sin(\theta)\sqrt{1-\gamma^2}}
\]
\[
\forall |z|\ge \gamma,\; \rho_{d,\gamma,\theta}(z)\le\frac{\sqrt{d}}{\sin(\theta)\sqrt{1-\gamma^2}}
\exp\left(-(d-1)\frac{(|z|-\gamma)^2}{4\sin^2(\theta)}\right)
\]
\end{lemma}
To prove lemma \ref{lem:bound_on_truncated_dist}, we will use an explicit formula for $\rho_{d,\gamma,\theta}$.
It will be convenient to introduce some notation. Let $\rho_{d,r}:\mathbb{R}\to \mathbb{R}$ be the density function of the random variable that is the inner product of a fixed unit vector in $S^{d-1}$ and a uniform vector in $r\cdot S^{d-1}$. Clearly,
\begin{equation}\label{eq:f_d_r_using_f_d_1}
\rho_{d,r}(x)= \frac{1}{r}\cdot \rho_{d,1}\left(\frac{x}{r}\right)
\end{equation}
We will use the following well known inequality
\begin{equation}\label{eq:sphere_ineq}
\rho_{d}(x)\le \sqrt{d}\exp\left(-\frac{x^2 d}{4}\right)
\end{equation}

\begin{lemma}\label{lem:expression_for_rho}
Let $A$ be the probability of $T_{d,\gamma}(w)$ according to the uniform distribution. We have
\[
\rho_{d,\gamma,\theta}(z) = \frac{1}{A}\int_{-\gamma\cos(\theta)}^{\gamma\cos(\theta)}
\rho_{d,\cos (\theta)}\left(u\right)\cdot
\rho_{d-1,\sqrt{\sin^2(\theta)-\tan^2(\theta)u^2}}\left(z-u\right)du
\]
\end{lemma}
\proof
Let $x$ be a uniform vector in the strip \mbox{$T_{d,\gamma}(w)$}, and let $y=\inner{w^*,x}$. We note that $\rho_{d,\gamma,\theta}$ is the density of $y$. We write
\[
x=\alpha\cdot w + z
\]
where $\inner{w,z}=0$. For $(w^*)^{\perp}=w^*-\inner{w^*,w}w$ we have,
\begin{eqnarray*}
y=\inner{w^*,x} &=& \alpha\cdot \inner{w^*,w} + \inner{w^*,z}
\\
&=& \alpha\cdot \cos(\theta) + \inner{(w^*)^{\perp},z}
\end{eqnarray*}
We note that the density function of the distribution of $\alpha\cdot \cos(\theta)$ is given by
\[
\tau(u)=\begin{cases}
\frac{1}{A}\rho_{d,\cos(\theta)}(u) & |u|\le \gamma\cdot\cos(\theta)
\\
0 & |u|>\gamma\cdot\cos(\theta)
\end{cases}
\]
Now, given $\alpha$, $z$ is a uniform vector of norm $\sqrt{1-\alpha^2}$ in the orthogonal complement of $w$, and $(w^*)^{\perp}$ is a vector of norm $\sin(\theta)$ in that space. It follows that the density function of $\inner{(w^*)^{\perp},z}$ given that $\alpha\cdot\cos(\theta)=u$ is \mbox{$\rho_{d-1,\sin(\theta)\cdot\sqrt{1-\frac{u^2}{\cos^2(\theta)}}}=
\rho_{d-1,\sqrt{\sin^2(\theta)-\tan^2(\theta)u^2}}$}. It therefore follows that
\[
\rho_{d,\gamma,\theta}(z) = \frac{1}{A}\int_{-\gamma\cos(\theta)}^{\gamma\cos(\theta)}
\rho_{d,\cos (\theta)}\left(u\right)\cdot
\rho_{d-1,\sqrt{\sin^2(\theta)-\tan^2(\theta)u^2}}\left(z-u\right)du
\]
\proofbox
We are now ready to prove lemma \ref{lem:bound_on_truncated_dist}.

\proof (of lemma \ref{lem:bound_on_truncated_dist})
Let $A$ be the probability of the strip $T_{d,\gamma}(w)$ according to the uniform distribution on the sphere. We have, using equations (\ref{eq:f_d_r_using_f_d_1}) and (\ref{eq:sphere_ineq}),
\begin{eqnarray*}
\rho_{d,\gamma,\theta}(z)&=&\frac{1}{A}\int_{-\gamma\cos(\theta)}^{\gamma\cos(\theta)}
\rho_{d,\cos (\theta)}\left(u\right)\cdot
\rho_{d-1,\sqrt{\sin^2(\theta)-\tan^2(\theta)u^2}}\left(z-u\right)du
\\
&\le & \frac{1}{A}\int_{-\gamma\cos(\theta)}^{\gamma\cos(\theta)}
\rho_{d,\cos (\theta)}\left(u\right)\cdot
\rho_{d-1,\sqrt{\sin^2(\theta)-\tan^2(\theta)u^2}}\left(0\right)du
\\
&\le & 
\frac{\sqrt{d-1}}{\sin(\theta)\sqrt{1-\gamma^2}}
\end{eqnarray*}
Similarly, for $|z|>\gamma$,
\begin{eqnarray*}
\rho_{d,\gamma,\theta}(z)&=&\frac{1}{A}\int_{-\gamma\cos(\theta)}^{\gamma\cos(\theta)}
\rho_{d,\cos (\theta)}\left(u\right)\cdot
\rho_{d-1,\sqrt{\sin^2(\theta)-\tan^2(\theta)u^2}}\left(z-u\right)du
\\
&\le & \frac{1}{A}\int_{-\gamma\cos(\theta)}^{\gamma\cos(\theta)}
\rho_{d,\cos (\theta)}\left(u\right)\cdot
\rho_{d-1,\sqrt{\sin^2(\theta)-\tan^2(\theta)u^2}}\left(|z|-\gamma\right)du
\\
&\le & \frac{1}{A\sin(\theta)\sqrt{1-\gamma^2}}\int_{-\gamma\cos(\theta)}^{\gamma\cos(\theta)}
\rho_{d,\cos (\theta)}\left(u\right)\cdot
\rho_{d-1,1}\left(\frac{|z|-\gamma}{\sqrt{\sin^2(\theta)-\tan^2(\theta)u^2}}\right)du
\\
&\le & \frac{1}{A\sin(\theta)\sqrt{1-\gamma^2}}\int_{-\gamma\cos(\theta)}^{\gamma\cos(\theta)}
\rho_{d,\cos (\theta)}\left(u\right)\cdot
\rho_{d-1,1}\left(\frac{|z|-\gamma}{\sin(\theta)}\right)du
\\
&= & \frac{1}{\sin(\theta)\sqrt{1-\gamma^2}}
\rho_{d-1,1}\left(\frac{|z|-\gamma}{\sin(\theta)}\right)
\\
&\le & \frac{\sqrt{d}}{\sin(\theta)\sqrt{1-\gamma^2}}
\exp\left(-(d-1)\frac{(|z|-\gamma)^2}{4\sin^2(\theta)}\right)
\end{eqnarray*}
\proofbox

\proof (of lemma \ref{lem:halfspace_appr_no_restriction})
By equation (\ref{eq:appr_half_by_appr_sign}), in is enough to show that the there is a univariate polynomial $p$ of degree $r=O\left(\frac{\log^2\left(1/\tau\right)}{\tau^2}\right)$ such that
\[
\int_{-1}^1 |p(x)-\sign(x)|\rho_{d,1}(x)dx\le \tau~.
\]
This, however, follows immediately from lemma \ref{lem:sig_appr_sub_gauss} and equation (\ref{eq:sphere_ineq}).
\proofbox

\paragraph{Acknowledgements:}
Amit Daniely is a recipient of the Google Europe Fellowship in Learning Theory, and this research is supported in part by this Google Fellowship. The author thanks Pranjal Awasthi, Adam Klivans, Nati Linial, and Shai Shalev-Shwartz for valuable discussions and comments.
\bibliography{bib}

\end{document}